\documentclass[12pt]{article}   
\usepackage{amsmath,graphicx}
\usepackage{amssymb, amsxtra} 
\usepackage{blkarray}
\usepackage{mathtools}

\usepackage{graphicx}
\usepackage{yfonts}
\def\therefore{
\leavevmode
\lower0.1ex\hbox{$\bullet$}
\kern-0.2em\raise0.9ex\hbox{$\bullet$}
\kern-0.2em\lower0.2ex\hbox{$\bullet$}
\thinspace}

\title{Quark mass matrices inspired by a numerical relation}
\author{A. Kleppe\footnote{astri.kleppe@gmail.com}\\ SACT, Oslo}
\date{}
\begin{document}
\maketitle

\begin{abstract}
\noindent In 1981, Yoshio Koide noticed that the square root values of the charged lepton masses satisfy the relation
\[ 
Q=\frac{m_e+m_{\mu}+m_{\tau}}{(\sqrt{m_e}+\sqrt{m_{\mu}}+\sqrt{m_{\tau}})^2} \approx \frac{2}{3} 
\]
Inspired by this relation, we introduce tentative mass matrices, using numerical values, and find matrices that display an underlying democratic texture.   
\end{abstract}

\section{Introduction}
A numerical relation involving the square roots of the lepton masses have intrigued people ever since it was published by Yoshio Koide in the beginning of the 1980-ies\cite{koide}.

\noindent It may be a purely coincidental relation, but we nevertheless use it as a source of inspiration, 
looking for mass matrices with a form that somehow corresponds to the Koide formula, using the square roots of the mass values. From the matrix ans\"{a}tze for the square roots of particle masses, we then derive mass matrices for up- and down-quarks, and investigate how the two charge sectors are related.   

Using the mass values $m_e=0.510998946 \hspace{1mm}MeV$, $m_{\mu}=105.6583745 \hspace{1mm}MeV$, and $m_{\tau}=1776.86 \hspace{1mm}MeV$,
Koide discovered that the square root values of the charged lepton masses satisfy the relation
\begin{equation}\label{koide} 
Q=\frac{m_e+m_{\mu}+m_{\tau}}{(\sqrt{m_e}+\sqrt{m_{\mu}}+\sqrt{m_{\tau}})^2} = 0.6666617\approx \frac{2}{3} 
\end{equation}
This is tantalizing, since it seems to echo the neat rational quantum numbers, like e.g. the electric charges of the elementary particles.
Many attempts have been made to interpret the relation (\ref{koide}), hoping to squeeze some insight out of it.
Denoting the square roots of the lepton masses as 
\begin{align*}
m_1=&\sqrt{m_e}\\ 
m_2=&\sqrt{m_{\mu}}\\ 
m_3=&\sqrt{m_{\tau}}, 
\end{align*}
and taking $T=m_1+m_2+m_3$ as a fixed number, we can interpret the set $(m_1,m_2,m_3)$ as a $partition$ of the nu
mber $T$, which in principle can be expressed as a sum of any three numbers, where the extreme cases are $T=A+A+A$ and $T=0+0+T$.

The numbers $m_1,m_2,m_3$ can also be perceived as the eigenvalues of a 3$\times$3 matrix $M$, and if we compare the  
quantity $Q=Trace(M^2)/(Trace(M))^2$ for the three cases
\[
T=A+A+A, \hspace{4mm}T=m_1+m_2+m_3, \hspace{4mm}T=0+0+T,
\]
we see that 
\begin{align*}
Q_{(A,A,A)}&=\frac{3A^2}{(3A)^2}=\frac{1}{3}\\
Q_{lepton}&=\frac{2}{3}\\
Q_{(0,0,T)}&=\frac{T^2}{(0+0+T)^2}=1,
\end{align*}
so the partition corresonding to the charged lepton sector lies right between the two extremes $(A,A,A)$ and $(0,0,T)$, which might be interpreted as the lepton mass spectrum having maximal amount of ``structure''.
\begin{figure}[htb]
    \begin{center}
    \includegraphics[scale=0.87]{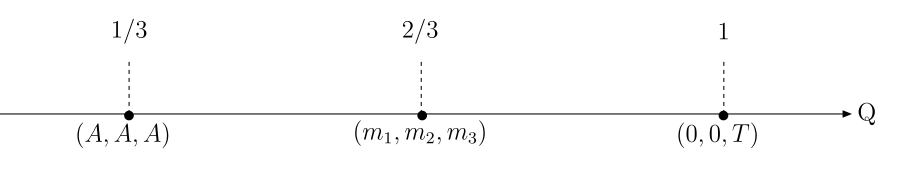}
\end{center}
\end{figure}
\noindent
\newline It should be noted that for the square roots of the running charged lepton masses at ${\bf{M}}_Z$ around 91 GeV, the results no longer give the exact Koide formula. The relation is however still of interest, because the Koide relation suggests that investigating the square roots of particle masses works as a kind of scale compression. This lessens the overwhelming impact of the largest masses, which tend to make the smallest masses irrelevant.

\section{Matrix invariants}
A 3$\times$3 matrix $M$ with eigenvalues $m_1,m_2,m_3$ has the invariants 
\begin{equation}
  \def\arraystretch{1.1}
  \begin{array}{r@{\;}l} 
Trace(M) =  & m_1+m_2+m_3\\

C_2(M) = &m_1m_2+m_1m_3+m_2m_3 \\

Det(M) = &m_1m_2m_3
  \end{array}
\end{equation}
Now let $\sqrt{M}_l$ be a matrix whose eigenvalues are the weighted square roots of the lepton masses, $(x_1,x_2,x_3)=(\sqrt{m_e},\sqrt{m_{\mu}},\sqrt{m_{\tau}})/N$, where $N=Trace(\sqrt{M}_l)/3$. We can then express the matrix invariants for the charged leptons as
\begin{align*}\label{leptons}
Trace(\sqrt{M}_l)  &= \sqrt{m_e}+\sqrt{m_{\mu}}+\sqrt{m_{\tau}}= 3N\\ 
C_2(\sqrt{M}_l)   &=\sqrt{m_e}\sqrt{m_{\mu}}+\sqrt{m_e}\sqrt{m_{\tau}}+\sqrt{m_{\mu}}\sqrt{m_{\tau}} \approx \frac{3}{2}N^2\\ 
Det(\sqrt{M}_l)  &= \sqrt{m_e}\sqrt{m_{\mu}}\sqrt{m_{\tau}} \approx \frac{N^3}{18}\\ 
\end{align*}
Again comparing the matrix invariants for the observed lepton mass spectra, with the invariants for the extreme spectra $(0,0,T)$ and $(A,A,A)$, we get
\begin{align*}
&Trace(0,0,T)  = T, \hspace{4mm}Trace(A)  = 3A,\hspace{3mm} \text{and}\hspace{3mm}Trace(\sqrt{M}_l)  = 3N\\ 
&C_2(0,0,T)    = 0, \hspace{10mm}C_2(A)    = 3A^2,\hspace{9mm} \text{and}\hspace{3mm}C_2(\sqrt{M}_l)    \approx \frac{3}{2}N^2\\ 
&Det(0,0,T)    = 0, \hspace{7mm}Det(A)    = A^3,\hspace{10mm} \text{and}\hspace{3mm}Det(\sqrt{M}_l)    \approx \frac{N^3}{18}\\ 
\end{align*}
and we again see that the invariants for the charged lepton sector lie between the two extremes.

\section{Mass states and flavour states}
When we talk about mass matrices, it is the form, or texture, of the mass matrices that we are looking for, in the hope to find a clue to the mechanism behind the hierarchical fermion mass spectra.

The mass matrices whose form we want to investigate, appear in the mass Lagrangian
${\mathcal{L}}_{mass}=\bar{\psi}M\psi$.
These mass matrices live in the weak basis, meaning that they are not in themselves measurable, but related to the measurable mass eigenstates by unitary rotation matrices $U$, 
\begin{equation}  
  {\mathcal{L}}_{mass}=\bar{\psi}M\psi=\bar{\psi}U^{\dagger}UMU^{\dagger}U\psi=\bar{\psi}_{phys}D\psi_{phys}
\end{equation}
where $\psi$ and $\psi_{phys}=U\psi$ denote the flavour states and the physical states, respectively, and $D=diag(m_1,m_2,m_3)$ is the diagonal mass eigenmatrix containing the masses of the physical particles of a given charge sector. 
\begin{figure}[htb]
    \begin{center}
    \includegraphics[scale=0.87]{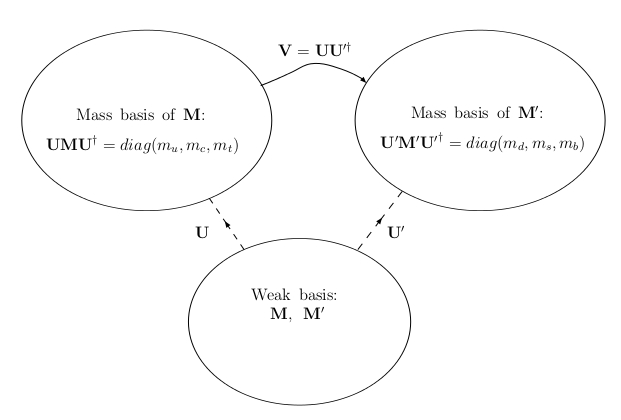}
\end{center}
\end{figure}
\newline
Our picture that massive up-quarks and massive down-quarks live in different mass bases, is based on the experimental fact that the Cabbibo-Kobayashi-Maskawa (CKM) mixing matrix $V_{CKM}$\cite{CKM} connecting the mass basis of the up-quarks with the mass basis of the down-quarks, deviates from the unit matrix. The mixing matrix appears in the charged current Lagrangian
\begin{equation}\label{cc}
{\mathcal{L}}_{cc}=-\frac{g}{2\sqrt{2}}\bar{\psi}_L\gamma^{\mu}V\psi'_LW_{\mu} + h.c. 
\end{equation}
where $\psi$ and $\psi'$ are fermion fields with charges $Q$ and $Q-1$, correspondingly.
  \newline 
It can be argued that flavour states merely exist in our imagination, since they are not directly measurable. This line of thought is however defied by the neutrinos, which as far as we know always appear as flavour states. Neutrino $\it{mass}$ $\it{states}$, on the other hand, never appear on the scene - in the sense that they never take part in interactions, but merely propagate in free space. The observed neutrinos  
$\nu _{e}, \nu _{\mu }, \nu _{\tau }$ are flavour states, but we nontheless perceive them as ``physical'', because they are the only neutrinos that ever appear in interactions, i.e. they are the only neutrinos that we ``see''.

\section{Democratic mass matrices}
We can perceive $(A,A,A)$ and $(0,0,T)$ as mass eigenvalues of the unit matrix
\[
A\begin{pmatrix}
 1  & 0 & 0\\
 0  & 1 & 0\\
 0  & 0 & 1\\
\end{pmatrix}
\]
and the democratic matrix\cite{Fr}\cite{Demo}
\[
\frac{T}{3}\begin{pmatrix}
 1  & 1 & 1\\
 1  & 1 & 1\\
 1  & 1 & 1\\
\end{pmatrix},
\] 
respectively, and guess that a relevant mass matrix for the leptons would be somewhere in between these matrices.

The democratic matrix represents a situation where at a zeroth level, all the particles within a given charge sector have the same Yukawa couplings.  
The argument for this assumption is that in the Standard Model, all fermions get their masses from the Yukawa couplings via the Higgs mechanism, and
since the couplings to the gauge bosons of the strong, weak and 
electromagnetic interactions are identical for all the fermions in a given charge sector, it seems like a natural assumption that they should also have identical Yukawa couplings. 

In the weak basis the democratic matrix $M_0$ is totally flavour symmetric, in the sense that the (weak) flavours $\psi_i$ of a given charge are indistinguishible (``absolute democracy''). This is contrary to experimental data, but it is reasonable to assume that actual mass matrices that represent physical particles, have some kind of modified democratic texture, since the mass spectrum $(0,0,T)$ of the democratic matrix 
reflects the experimental situation with one very heavy and two much lighter fermions. But in order to get correct, non-zero masses, the initial democratic matrix must clearly be modified.
\noindent A first step towards a more realistic mass spectrum is to introduce an extra parameter while keeping the trace constant, for example like  
\begin{equation}
M_1 =
 N\begin{pmatrix}
 1  & 1 & X\\
 1  & 1 & X\\
 X  & X & 1\\
\end{pmatrix}
 \end{equation}
which has the mass spectrum $N(0,\frac{1}{2}(3-\sqrt{1+8X^2}),\frac{1}{2}(3+\sqrt{1+8X^2}))$.
A similar matrix,
\begin{equation}
M_2 =
 N\begin{pmatrix}
 1  & X & X\\
 X  & 1 & X\\
 X  & X & 1\\
\end{pmatrix}
 \end{equation}
has three non-zero eigenvalues, $N(1-X, 1-X, 2X+1)$, but two of them are degenerate.
\newline In order to obtain three physical, non-degenerate, non-zero masses, we introduce yet another parameter, and still keeping the trace constant, we write  
\begin{equation}
M_3 =
 N\begin{pmatrix}
 1  & Y & X\\
 Y  & 1 & X\\
 X  & X & 1\\
\end{pmatrix}
 \end{equation}   
which has the mass spectrum $N(1-Y,\frac{1}{2}(2+Y-\sqrt{8X^2+Y^2}),\frac{1}{2}(2+Y+\sqrt{8X^2+Y^2}))$.

\noindent Let us assume that this represents the (square roots of the masses of the) leptons, so $Trace(\sqrt{M}_l) = 3N_{lepton}$,
where $N= 17.716 \sqrt{\hspace{1mm}MeV}$, and $X$ and $Y$ are dimensionless coefficients.
The matrix invariants are
\[
C_2(\sqrt{M}_l)= N^2(3-Y^2-2X^2)\hspace{2mm}\text{and}\hspace{2mm}
Det(\sqrt{M}_l)=N^3(1-Y)(1+Y-2X^2),
\]
which for $(x_1,x_2,x_3)=(\sqrt{m_e},\sqrt{m_{\mu}},\sqrt{m_{\tau}})/N$, gives 
\begin{align*}
&x_1=(1-Y)\\ 
&x_2=(2+Y-\sqrt{Y^2+8X^2})/2\\ 
&x_3=(2+Y+\sqrt{Y^2+8X^2})/2,\\ 
  \end{align*}
i.e
\begin{align*}   
Y    &= 1-\sqrt{m_e}/N\\
2X^2 &=1+Y-\frac{\sqrt{m_{\mu}m_{\tau}}}{N^2}=2-\frac{\sqrt{m_e}}{N}-\frac{\sqrt{m_{\mu}m_{\tau}}}{N^2} \\
\end{align*}
With 
the lepton mass values $m_e=0.51099 \hspace{1mm}MeV$, $m_{\mu}=105.6584 \hspace{1mm}MeV$, $m_{\tau}=1776.86 \hspace{1mm}MeV$,
we get the numerical values for the coefficients
\begin{align*}
Y    &= 0.9597\\
X    &= 0.538\\
\end{align*}
\noindent  
Inserting these values into the matrix invariants, we get $C_2(\sqrt{M})=N^2\times 1.500027$ and 
$Det(\sqrt{M}_l)=N^3/17.95$, which is reasonably close to $C_2(\sqrt{M})= 3N^2/2 $ and
and $Det(\sqrt{M})= N^3/18$.

We can add complexity to the matrix, e.g. by multiplication with the matrix
\[
\begin{pmatrix}
 e^{i\alpha}  & 0 & 0\\
 0  & 1 & 0\\
 0  & 0 & 1\\
\end{pmatrix}
 \]
 and its conjugate, and obtain the final mass matrix
 \begin{equation}
 \sqrt{M}_l= N \begin{pmatrix}
     e^{i\alpha}  & 0 & 0\\
 0  & 1 & 0\\
 0  & 0 & 1\\
\end{pmatrix}
\begin{pmatrix}
 1  & Y & X\\
 Y  & 1 & X\\
 X  & X & 1\\
\end{pmatrix}
\begin{pmatrix}
 e^{-i\alpha}  & 0 & 0\\
 0  & 1 & 0\\
 0  & 0 & 1\\
\end{pmatrix}=
N\begin{pmatrix}
 1 & Ye^{i\alpha} & Xe^{i\alpha}\\
 Ye^{-i\alpha}  & 1 & X\\
X e^{-i\alpha}  & X & 1\\
\end{pmatrix}
\end{equation}

\section{Quarks}
The possibility that quark masses display a pattern similar to the Koide formula, has of course been examined by many authors \cite{koi}.
\newline We look for relations similar to (\ref{koide}) in the quark sector, and use these relations as a basis for new ans\"{a}tze of the mass matrices of the down- and up-setors, respectively, with the ultimate goal og get a notion of how the mass matrices are related in the weak basis. 

Here we use the following mass values for the up- and down sectors 
\cite{Jamin}, \cite{Jamin2}.
\begin{equation}\label{jamin} 
\begin{matrix}
m_u(M_z)=1.24 \hspace{1mm}MeV, & m_c(M_z)= 624 \hspace{1mm}MeV, & m_t(M_z) = 171550 \hspace{1mm}MeV\\
m_d(M_z)=2.69 \hspace{1mm}MeV, & m_s(M_z)= 53.8 \hspace{1mm}MeV, & m_b(M_z) = 2850 \hspace{1mm}MeV \\
\end{matrix} 
\end{equation}
and taking the square roots, 
we get
\begin{equation} 
\def\arraystretch{1.1}
  \begin{array}{r@{\;}l} 
Q_d     =\frac{m_d+m_s+m_b}{(\sqrt{m_d}+\sqrt{m_s}+\sqrt{m_b})^2}      \approx & 3/4\\
Q_u     =\frac{m_u+m_c+m_t}{(\sqrt{m_u}+\sqrt{m_c}+\sqrt{m_t})^2}       \sim & 8/9\\
\end{array}
\end{equation} 
For all the charged fermion sectors, we use the parametrization
$Tr(\sqrt{M})=KN$ where $M$ is a 3 $\times$ 3 matrix, and $K$ is an integer.
This gives us
\begin{equation}\label{D} 
\def\arraystretch{1.1}
  \begin{array}{r@{\;}l} 
Tr(\sqrt{M})_d=&4 N\\
2C_2(\sqrt{M})_d\approx &4N^2\\
Det(\sqrt{M})_d\approx & N^3/6\\
\end{array}
\end{equation} 
and
\begin{equation}\label{U}  
\def\arraystretch{1.1}
  \begin{array}{r@{\;}l} 
Tr(\sqrt{M})_u=&9N\\
2C_2(\sqrt{M})_u\sim &9N^2\\
Det(\sqrt{M})_u\sim & N^3/10\\
\end{array}
\end{equation} 
The quark masses are however not as well established as the lepton masses, which in addition are the only ones that satisfy an exact Koide relation.
\\
There are many possible choices for matrices with a given trace, 
we can for example have matrices of the form
\[
\sqrt{M}_d = 
 N_d \begin{pmatrix}
 1  & A & B\\
 A  & 1 & B\\
 B  & B & 2\\
\end{pmatrix},\hspace{3mm} 
\sqrt{M}_d' = 
 N_d \begin{pmatrix}
 0  & A & B\\
 A  & 1 & B\\
 B  & B & 3\\
 \end{pmatrix},\hspace{3mm}\text{or}\hspace{3mm} 
 \sqrt{M}_d'' = 
 N_d \begin{pmatrix}
 0  & A & B\\
 A  & 0 & B\\
 B  & B & 4\\
\end{pmatrix}
   \]    
   where $Trace(\sqrt{M}_d)=4N_d$, $N_d=(\sqrt{m_d}+\sqrt{m_s}+\sqrt{m_b})/4=15.59 \sqrt{\hspace{1mm}MeV}$, and $A$ and $B$ are dimensionless coefficients. With the democratic form as a guiding line, we choose to study the first matrix.

In the case of the up sector, we can for example study these matrices
\[
   {\sqrt{M}}_u =
N_u \begin{pmatrix}
1  & D &  E\\
 D  & 1 & E\\
 E  & E & 7\\
\end{pmatrix},\hspace{3mm}\text{or}\hspace{3mm}  {\sqrt{M}}_u' =
N_u \begin{pmatrix}
2  & D &  E\\
 D  & 2 & E\\
 E  & E & 5\\
\end{pmatrix}, \hspace{3mm}\text{or}\hspace{3mm}  {\sqrt{M}}_u'' =
N_u \begin{pmatrix}
3  & D &  E\\
 D  & 3 & E\\
 E  & E & 3\\
\end{pmatrix}
\]
\newline
Summing up: our point of departure is a mass matrix for the square roots of the down-quark masses, with a nearly democratic texture. Using numerical mass values and a numerical mixing matrix, we derive a matrix for (the square roots of) the up-quark masses
from our matrix ansatz for the down sector.

With a similar matrix ansatz for the up-quarks, we then derive a mass matrix for the down sector. This finally gives us two sets of quark mass matrices, which we study in order to find credible mass matrices for both charge sectors.

\section{Ans\"{a}tze for the down-quarks}
$\bf{1.}$
Our initial ansatz for the down sector (using the square roots of the down-quark masse), is the matrix 
\begin{equation}\label{down}
\sqrt{M}_d = 
 N_d \begin{pmatrix}
 1  & A & B\\
 A  & 1 & B\\
 B  & B & 2\\
\end{pmatrix}
 \end{equation}
with $Trace(\sqrt{M}_d)=4N_d$, $N_d=(\sqrt{m_d}+\sqrt{m_s}+\sqrt{m_b})/4=15.59 \sqrt{\hspace{1mm}MeV}$, and $A$ and $B$ are dimensionless coefficients. The matrix invariants read 
\[
\def\arraystretch{1.1}
\begin{array}{r@{\;}l}
&Tr(\sqrt{M}_d) =4 N_d \\
&2C_2(\sqrt{M}_d) = N_d^2(5-A^2-2B^2)\\
&Det(\sqrt{M}_d)= N_d^3(2+2AB^2-2A^2-2B^2)=2N_d^3(1-A)(1+A-B^2),\\
\end{array}
\]
and the dimensionless
eigenvalues are
\[
(x_1,x_2,x_3)=(1-A,\frac{1}{2}(A+3 -\sqrt{8B^2+(1-A)^2}),\frac{1}{2}(A+3 +\sqrt{8B^2+(1-A)^2})),
\]
from which we calculate the dimensionless coefficients 
\begin{align*}
A    &= 0.895\\
B    &= 1.044\\
\end{align*}
The diagonalizing matrix for the matrix (\ref{down}) is
\[
U_d=\begin{pmatrix}
\frac{1}{\sqrt{2}}                          &   -\frac{1}{\sqrt{2}}                   & 0\\
\frac{1}{2}\sqrt{\frac{1-A+S}{S}}     &\frac{1}{2}\sqrt{\frac{1-A+S}{S}}  &-\frac{2B}{\sqrt{S(1-A+S)}}\\
\frac{1}{2}\sqrt{\frac{A-1+S}{S}}     &\frac{1}{2}\sqrt{\frac{A-1+S}{S}}  &\frac{2B}{\sqrt{S(A-1+S)}}\\
\end{pmatrix}
\]
where $S=\sqrt{8B^2+(1-A)^2}$.

\noindent Using the definition of the weak mixing matrix $V_{CKM}=U_uU_d^{\dagger}$, together with $U_d^{\dagger}U_d=1$, we get 
\[
U_u=V_{CKM}U_d
\]
and using the numerical mixing matrix \cite{Mix}
\begin{equation}
V_{CKM}=\begin{pmatrix}
0.9737        & 0.2243       & 0.0038\\
0.221          & 0.975        & 0.0408\\  
0.0086         & 0.0415       & 1.014\\
\end{pmatrix}
\end{equation}
together with the numerical expression of $U_d$ obtained by
inserting the numerical values $A=0.895$ and $B=1.044$,
we get a numerical expression for $U_u$. This allows us to numerically calculate the mass matrix of the up sector.

We are operating at a completely phenomenological level, no theory, just investigating what diagonalization matrix for the up sector that comes together with the diagonalization matrix for the down sector, when we use the ansatz (\ref{down}) for the down sector.
\newline So the diagonal mass matrices for the up-quarks and the down-quarks, respectively, are tied together by the mixing matrix.
The numerical value of the (square root) mass matrix for the up sector is
\[
\sqrt{M}_u^{(derived)}=N_uU_u^{\dagger}diag(x_1,x_2,x_3)U_u,
\]
where $(x_1,x_2,x_3)=(\sqrt{m_u},\sqrt{m_c},\sqrt{m_t})/N_u$ and $N_u=(\sqrt{m_u}+\sqrt{m_c}+\sqrt{m_t})/9\sqrt{\hspace{1mm}MeV}$.
\newline With the numerical values $(x_1,x_2,x_3)=(0.03,0.509,8.461)\sqrt{\hspace{1mm}MeV}$, the derive the numerical mass matrix for the (square roots) of the up sector, 
\[
\sqrt{M}_u^{(derived)}=N_u\begin{pmatrix}
2.21  &   2.08  &  2.65\\
2.26  &  2.14  &  2.49\\
3.03  &  3.07 &   4.65\\
\end{pmatrix},
\]
For a generic matrix $\sqrt{M}$ with eigenvalues $m_i$, we have
$\sqrt{M}=U^{\dagger}diag(\sqrt{m_1},\sqrt{m_2},\sqrt{m_3})U$,
where $U$ and $U^{\dagger}$ diagonalize $\sqrt{M}$, and the final, non-square-root matrix is ${\bf{M}}=(\sqrt{M})^2$:
\[
(\sqrt{M})^2=(U^{\dagger}diag(\sqrt{m_1},\sqrt{m_2},\sqrt{m_3})U)^2=
U^{\dagger}diag(m_1,m_2,m_3) U = {\bf{M}},
\]
So the ``normal'' derived mass matrix for the up sector is 
\begin{equation}\label{dup}
{\bf{M}}_u^{(derived)}=N^2_u\begin{pmatrix}
17.64 &  17.20 &  23.4\\
17.38 &  16.93 & 22.92\\
27.70 &  27.12 &  37.28\\
\end{pmatrix}
\end{equation}
where $N^2_u$ has dimension $MeV$.

Inserting numerical values for $A$ and $B$ in the down sector mass matrix (\ref{down}), we get
\begin{equation}
{\bf{M}}_d^{(ansatz)}=N^2_d\begin{pmatrix}
2.89   &  2.88 &   4.07\\
2.88  & 2.89   &  4.07\\
4.07 & 4.07 &  6.18\\
\end{pmatrix},
\end{equation}
and when we rescale ${\bf{M}}_u^{(derived)}$, we see that the matrices have a similar texture:
\begin{equation}\label{d2}
{\bf{M}}_u^{(derived)}=5.97 \times N^2_u\begin{pmatrix}
2.95  & 2.88  & 3.92 \\
2.90   & 2.83  &  3.84\\
4.64  &4.54  &   6.24\\
\end{pmatrix},
\hspace{2mm}
{\bf{M}}_d^{(ansatz)}=N^2_d\begin{pmatrix}
2.89   &  2.88 &   4.07\\
2.88  & 2.89   &  4.07\\
4.07 & 4.07 &  6.18\\
\end{pmatrix}
\end{equation}

\subsection{Next ansatz}
$\bf{2.}$
 We now consider another down-quark matrix ansatz, 
\begin{equation}\label{6d}
\sqrt{M}_d=N_d\begin{pmatrix}
2 & A & B\\
A & 2 & B\\
B & B & 2\\
\end{pmatrix}
\end{equation}
where $N_d=(\sqrt{m_d}+\sqrt{m_s}+\sqrt{m_b})/6 =10.4 \sqrt{MeV}$.
From the matrix invariants
\[
C_2({\sqrt{M}}_d )=N_d^2(12-A^2-2B^2) \hspace{2mm}\text{and}\hspace{2mm}
Det({\sqrt{M}}_d )=N_d^32(2-A)(2+A-B^2)
\]
we get the dimensionless eigenvalues
\[
(x_1,x_2,x_3)=(2-A, \frac{1}{2}(A+4 -\sqrt{A^2+8B^2}),\frac{1}{2B}(A+4 +\sqrt{A^2+8B^2}))
\]
Inserting numerical values from (\ref{jamin}), we get
\begin{align*}
A    &= 1.842\\
B    &= 1.425\\
N_d  &=10.4 \sqrt{\hspace{1mm}MeV}\\ 
\end{align*}
Inserting the numerical values for $A$ and $B$ in $U_d$, and using $U_u=V_{CKM}U_d$, we find the numerical expression for $\sqrt{M_u}=U_u^{\dagger}diag(\sqrt{m_u},\sqrt{m_c},\sqrt{m_t})U_u$, and 
we get the matrices 
\begin{equation}\label{222N}
{\bf{M_u}}^{(derived)}=N_u^2\begin{pmatrix}
25.8  &  25.28 &  21.1\\
25.47  &  24.93 &  20.72\\
25.37   & 24.95 &  21.12\\
\end{pmatrix}
\hspace{1mm}\text{and}\hspace{2mm}
{\bf{M}}_d^{(ansatz)}= N_d^2\begin{pmatrix}
9.426 &  9.4 & 8.325\\
9.4   &  9.426 &  8.325\\
8.325 & 8.325 & 8.06 \\                                        
\end{pmatrix}
\end{equation}
Rescaling the up-quark matrix:
\begin{equation}\label{222M}
{\bf{M_u}}^{(derived)}=2.74 \times N_u^2\begin{pmatrix}
9.43  &  9.24 &  7.71\\
9.3  &  9.11 &   7.6\\
9,42   & 9.12 &  7.72\\
\end{pmatrix}
\hspace{1mm}\text{and}\hspace{2mm}
{\bf{M}}_d^{(ansatz)}= N_d^2\begin{pmatrix}
9.43 &  9.4 & 8.33\\
9.4   &  9.43 &  8.33\\
8.33 & 8.33 & 8.06 \\                                        
\end{pmatrix}
\end{equation}
shows that the matrices for the up-sector and the down-sector have similar, nearly democratic structures.

\section{Ans\"{a}tze for the up-quarks}
We can play the game the other way round, by introducing a matrix ansatz for the square roots of the up-quark masses, from which we derive a matrix for the down sector.
\newline $\bf{1.}$ We first consider the matrix
\begin{equation}\label{up}
{\sqrt{M}}_u =
N_u \begin{pmatrix}
2  & D &  E\\
 D  & 2 & E\\
 E  & E & 5\\
\end{pmatrix}
 \end{equation}
where $N_u=(\sqrt{m_u}+\sqrt{m_c}+\sqrt{m_t})/9=49.09 \sqrt{\hspace{1mm}MeV}$, and $D$ and $E$ are dimensionless coefficients.
From the matrix invariants 
\[  
\def\arraystretch{1.1}
\begin{array}{r@{\;}l}
&Tr({\sqrt{M}}_u) =9N_u\\
&2C_2({\sqrt{M}}_u) =N_u^2(24-D^2-2E^2)\\
&Det({\sqrt{M}}_u)=N_u^3(20+2DE^2-5D^2-4E^2)=N_u^3(2-D)(5(1+D)-2E^2)\\
\end{array}
\]
we get that $x_ 1=(2-D)$, and the dimensionless 
eigenvalues are
\[
(x_1,x_2,x_3)=(2-D,\frac{1}{2}(D+7 -\sqrt{8E^2+(3-D)^2}),\frac{1}{2}(D+7 +\sqrt{8E^2+(3-D)^2}))
\]
where
\begin{align}
&D  = 1.97\\ 
&E   = 2.79\\ 
&N_u =49.09 \sqrt{\hspace{1mm}MeV}\\ 
\end{align}
Following a similar procedure as before,
using $U_d=V_{CKM}^{\dagger}U_u$,
using the diagonalization matrix
\begin{equation}
  U_u= \begin{pmatrix}
\frac{1}{\sqrt{2}}          & -\frac{1}{\sqrt{2}}                  & 0\\
\frac{1}{2}\sqrt{\frac{S-D+3}{S}} & \frac{1}{2}\sqrt{\frac{S-D+3}{S}} &  -\frac{2E}{\sqrt{S(S-D+3)}}\\
\frac{1}{2}\sqrt{\frac{S+D-3}{S}} & \frac{1}{2}\sqrt{\frac{S+D-3}{S}} &  \frac{2E}{\sqrt{S(S+D-3)}}\\
\end{pmatrix}
 \end{equation}
and again inserting the mass matrix for the square roots of the weighted mass values,$(x_1,x_2,x_3)= (\sqrt{m_d},\sqrt{m_s},\sqrt{m_b}/N_d$, with $N_d=(\sqrt{m_d}+\sqrt{m_s}+\sqrt{m_b})/4$, we get the mass matrix for the square roots of the down-quarks in the weak basis:
\begin{equation}\label{downderived}
\sqrt{M}_d^{(derived)}=N_dU_d^{\dagger}\begin{pmatrix}
0.105                & 0             & 0\\
0                 & 0.47             & 0\\
0                  & 0       &3.425\\
\end{pmatrix}U_d=
N_d\begin{pmatrix}
0.96 &  0.77 &  0.98\\
0.92 &  0.91 & 0.92\\
1.07 &  1.17 &  2.14\\
\end{pmatrix},
\end{equation}
which gives
\begin{equation}
{\bf{M}}_d^{(derived)}=N_d^2\begin{pmatrix}
2.67 &  2.58 & 3.73 \\
2.7 &  2.61 & 3.69 \\
4.39 &  4.38 & 6.68 \\
\end{pmatrix}
\hspace{1mm}\text{and}\hspace{2mm}
{\bf{M}}_u^{(ansatz)}=6 \times N_u^2\begin{pmatrix}
2.582 & 2.583 & 4.17\\
2.583 & 2.582 & 4.17\\
4.13  & 4.17  & 6.76\\ 
\end{pmatrix},
\end{equation}
which can be related to (\ref{d2}).

\subsection{Next ansatz}
{\bf{2.}}
We now consider an alternative ansatz for the up sector,
\begin{equation}\label{Up} 
{\sqrt{M}}_u = 
N_u \begin{pmatrix}
3  & D &  E\\
 D  & 3 & E\\
 E  & E & 3\\
\end{pmatrix}
\end{equation}
where $Trace({\sqrt{M}}_u)=9N_u$, $N_u=(\sqrt{m_u}+\sqrt{m_c}+\sqrt{m_t})/9=49.09 \sqrt{\hspace{1mm}MeV}$, and $D$ and $E$ are dimensionless coefficients.
The matrix invariants are
\[
C_2({\sqrt{M}}_u )=N_u^2(27-D^2-2E^2) \hspace{1mm}\text{and}\hspace{1mm}
Det({\sqrt{M}}_u )=N_u^3(3-D)(9+3D-2E^2)
\]
which gives the dimensionless eigenvalues
\[
(x_1,x_2,x_3)=(3-D, \frac{1}{2}(D+6 -\sqrt{D^2+8E^2}),\frac{1}{2}(D+6 +\sqrt{D^2+8E^2}))
\]
Inserting numerical values from (\ref{jamin}), we get
\begin{align*}
D    &= 2.97\\
E    &= 2.61\\
N_u  &=49.09 \sqrt{\hspace{1mm}MeV}\\ 
\end{align*}
Following the same procedure as above, we get
the matrix
\[
{\bf{M}}_d^{(derived)}=N_d^2\begin{pmatrix}
4.08  &  4.02 &  3.48\\
4.10 &  4.04 &  3.42\\
4.11 &  4.14  &  3.84\\
\end{pmatrix}
\]
which we compare to
\begin{equation}\label{333}
{\bf{M}}_u^{(ansatz)}=N_u^2\begin{pmatrix}
24.623 &     24.622 &      23.39\\
24.622  &      24.623 &  23.39\\
23.39  &     23.39  &     22.6\\
\end{pmatrix}
\end{equation}
The mass matrix from the ansatz and the derived matrix are again of similar texture, which is close to democratic.

\noindent If we instead use the weighting $N_d=(\sqrt{m_d}+\sqrt{m_s}+\sqrt{m_b})/6$ for the down sector,
we get
\begin{equation}\label{6}
{\bf{M}}_u^{(ansatz)}=2.68 \times N_u^2\begin{pmatrix}
9.1886 &     9.1882 &      8.73\\ 
9.1882  &    9.1886 &  8.73\\
8.73  &   8.73  &    8.44\\
\end{pmatrix}
\hspace{2mm}\text{and}\hspace{2mm}
{\bf{M}}(6)_d^{(derived)}=N_d(6)^2\begin{pmatrix}
9.18 &  9.04 &  7.83\\
9.23  &   9.09 &  7.69\\
9.25 &   9.31  & 8.64\\
\end{pmatrix},
\end{equation}
which reflects the matrices (\ref{222M}).

\noindent The above procedures can be repeated with the quark masses $m_q(2 GeV)$\cite{PDG},
leading to similar results.

\section{Discussion}
The ansatz for the square roots of the up-quark masses
\[
{\sqrt{M}}_u(3,3,3) = 
N_u \begin{pmatrix}
3  & D &  E\\
 D  & 3 & E\\
 E  & E & 3\\
\end{pmatrix}
\]
with $D=2.97$ and $E=6.08$, gives the regular up-quark matrix 
\[
{\bf{M}}_u^{(ansatz)}=N_u^2\begin{pmatrix}
24.623 &     24.622 &      23.39\\
24.622  &      24.623 &  23.39\\
23.39  &     23.39  &     22.6\\
\end{pmatrix},
\]
which is similar to the up-quark matrix that we derive from the matrix
\[
\sqrt{M}_d(2,2,2)=N_d\begin{pmatrix}
2 & A & B\\
A & 2 & B\\
B & B & 2\\
\end{pmatrix}
\]   
which, with $A=1.843$ and $B=1.425$ inserted, is
\[
{\bf{M_u}}^{(derived)}=N_u^2\begin{pmatrix}
25.8  &  25.28 &  21.1\\
25.47  &  24.93 &  20.72\\
25.37   & 24.95 &  21.12\\
\end{pmatrix}
\]
We therefore perceive that the pair of matrices ${\sqrt{M}}_u(3,3,3)$ and ${\sqrt{M}}_d(2,2,2)$ as belonging together, in the sense that from a matrix of the form $\sim {\sqrt{M}}_u(3,3,3)$ for the up-quarks, we derive a matrix for the down-quarks of the form $\sim {\sqrt{M}}_u(2,2,2)$ - and vice versa.
\newline
In the same way, we perceive the matrices
\[
\sqrt{M}_d(1,1,2) = 
 N_d \begin{pmatrix}
 1  & A & B\\
 A  & 1 & B\\
 B  & B & 2\\
\end{pmatrix}
 \]
and
\[
 {\sqrt{M}}_u(2,2,5) =
N_u \begin{pmatrix}
2  & D &  E\\
 D  & 2 & E\\
 E  & E & 5\\
\end{pmatrix},
\]
as belonging together.

\noindent Higher powers of this type of matrices, with diagonal $(X,X,Y)$ will asymptotically go towards matrices with the texture 
\[
{\sqrt{M}} \sim
N \begin{pmatrix}
F  & F-\epsilon &  G\\
F-\epsilon  & F & G\\
G  & G & H\\
\end{pmatrix}
 \]
which have a mass spectrum of the type $(\epsilon, \frac{1}{2}(H+2F-\epsilon \pm \sqrt{(H+\epsilon-2F)^2+8G^2})))$.
\noindent
Moreover, the (square root) quark mass matrices of the type 
\[
{\sqrt{M}} \sim
N \begin{pmatrix}
 H & D &  E\\
 D  & H & E\\
 E  & E & H\\
\end{pmatrix},
\]
with eigenvalues $(H-2,\frac{1}{2}(D+2H \pm \sqrt{D+8E^2}))$,
give rise to matrices with a more democratic texture.

\section{Appendix}
$\bullet$ The diagonalizing matrix for the matrix (\ref{down})
\[
{\sqrt{M}}_d =
N_d \begin{pmatrix}
1  & A &  B\\
A  & 1 & B\\
 B  & B & 2\\
\end{pmatrix}
\]
is
\[
U_d=\begin{pmatrix}
\frac{1}{\sqrt{2}}                          &   -\frac{1}{\sqrt{2}}                   & 0\\
\frac{1}{2}\sqrt{\frac{1-A+S}{S}}     &\frac{1}{2}\sqrt{\frac{1-A+S}{S}}  &-\frac{2B}{\sqrt{S(1-A+S)}}\\
\frac{1}{2}\sqrt{\frac{A-1+S}{S}}     &\frac{1}{2}\sqrt{\frac{A-1+S}{S}}  &\frac{2B}{\sqrt{S(A-1+S)}}\\
\end{pmatrix}
\]
where $S=\sqrt{8B^2+(1-A)^2}$.

$\bullet$ The diagonalizing matrix for the matrix (\ref{6d}) 
\[
{\sqrt{M}}_d =
N_d \begin{pmatrix}
2  & A &  B\\
A  & 2 & B\\
 B  & B & 2\\
\end{pmatrix}
\]
is
\[
  U_d= \begin{pmatrix}
    \frac{1}{\sqrt{2}}      & -\frac{1}{\sqrt{2}}  & 0\\
   \frac{2B}{\sqrt{2S(S+A)}}    & \frac{2B}{\sqrt{2S(S+A)}}   & -\sqrt{\frac{S+A}{2S}}\\
   \frac{2B}{\sqrt{2S(S-A)}} &  \frac{2B}{\sqrt{2S(S-A)}}  & \sqrt{\frac{S-A}{2S}}\\
\end{pmatrix}
  \]
where $S=\sqrt{8B^2+A^2}$.  

$\bullet$ The ansatz (\ref{up}) for the up sector:
\[
{\sqrt{M}}_u =
N_u \begin{pmatrix}
2  & D &  E\\
 D  & 2 & E\\
 E  & E & 5\\
\end{pmatrix}
\]
has the diagonalization matrix
\begin{equation}
  U_u= \begin{pmatrix}
\frac{1}{\sqrt{2}}          & -\frac{1}{\sqrt{2}}                  & 0\\
\frac{1}{2}\sqrt{\frac{S-D+3}{S}} & \frac{1}{2}\sqrt{\frac{S-D+3}{S}} &  -\frac{2E}{\sqrt{S(S-D+3)}}\\
\frac{1}{2}\sqrt{\frac{S+D-3}{S}} & \frac{1}{2}\sqrt{\frac{S+D-3}{S}} &  \frac{2E}{\sqrt{S(S+D-3)}}\\
\end{pmatrix}
 \end{equation}
 where $S=\sqrt{3-D)^2}$.  

$\bullet$ The ansatz (\ref{up}) for the up sector
\[
{\sqrt{M}}_u =
N_u \begin{pmatrix}
3  & D &  E\\
 D  & 3 & E\\
 E  & E & 3\\
\end{pmatrix}
\]
has the diagonaization matrix
\begin{equation}
  U_u= \begin{pmatrix}
\frac{1}{\sqrt{2}}          & -\frac{1}{\sqrt{2}}                  & 0\\
\frac{1}{2}\sqrt{\frac{S-D}{S}} & \frac{1}{2}\sqrt{\frac{S-D}{S}} &  -\frac{2E}{\sqrt{S(S-D)}}\\
\frac{1}{2}\sqrt{\frac{S+D}{S}} & \frac{1}{2}\sqrt{\frac{S+D}{S}} &  \frac{2E}{\sqrt{S(S+D)}}\\
\end{pmatrix}
 \end{equation}
where $S=\sqrt{D^2+8E^2}$.

\section{Conclusion}
Inspired by the Koide relation, we investigate two types of mass matrices for the square roots of the quark masses, from which we derive the regular mass matrices. 
Using the numerical Cabbibo-Kobayashi-Maskawa matrix $V_{CKM}$, we derive numerical up-quark matrices from numerical down-quark matrix ans\"{a}tze, and similarly, we derive numerical down-quark matrices from numerical up-quark matrix ans\"{a}tze. We are well aware of the uncertainty of the numerical quark mass values, but we believe that we still get an indication of (flavour space) relations between up-quark mass matrices and down-quark mass matrices.      

\noindent
The matrices we have investigated are of the type
\begin{equation}\label{K}
\sqrt{M}=N\begin{pmatrix}
  K  & A & B\\
  A  & K & B\\
  B  & B & P\\
\end{pmatrix}
\end{equation}
where $K,P$ are integers, and $N=(\sqrt{m_1}+\sqrt{m_2}+\sqrt{m_3})/(2K+P)$, where $m_j$ are quark masses, and
\begin{equation}\label{L}
\sqrt{M}=N\begin{pmatrix}
  K  & A & B\\
  A  & K & B\\
  B  & B & K\\
\end{pmatrix}
\end{equation}
where $K$ is an integer, and $N=(\sqrt{m_1}+\sqrt{m_2}+\sqrt{m_3})/(3K)$.
Starting from a matrix of the form (\ref{K}) for the down sector, the matrix that we numerically derive for the up sector, has a texture similar to (\ref{K}) (and vice versa starting from the up sector).
Likewise, starting from a matrix of the form (\ref{L}) for the down sector, we numerically derive a similar matrix for the up sector (and vice versa starting with an ansatz for the up sector).
The mass matrices derived from (\ref{L}) moreover have a nearly democratic structure. 

\noindent So the conclusion is the mass matrices for the two sectors tend to have similar textures, which is not surprising, since $V_{CKM}$ is not so terribly far from the unit matrix.

\end{document}